\documentclass[12pt,preprint]{aastex}
\usepackage{lscape}

\def\rel{{\rm rel}}
\def\e{{\rm E}}
\def\au{{\rm AU}} 
\def\muas{{\mu\rm as}}

\def\kpc{{\rm kpc}}
\def\pc{{\rm pc}}
\def\rel{{\rm rel}}

\def\mas{{\rm mas}}
\def\masyr{{\rm mas\,yr^{-1}}}

\def\e{{\rm E}}

\begin{document}
\title{Microlens OGLE-2005-BLG-169 Implies Cool Neptune-Like Planets 
are Common}

\author{
A.~Gould\altaffilmark{1,6},
A.~Udalski\altaffilmark{2,7},
D.~An\altaffilmark{1,6},
D.P.~Bennett\altaffilmark{4,5,8},
A.-Y.~Zhou\altaffilmark{9},
S.~Dong\altaffilmark{1,6},
N.J.~Rattenbury\altaffilmark{4,10},
B.S.~Gaudi\altaffilmark{1,11},
P.C.M.~Yock\altaffilmark{4,12},
I.A.~Bond\altaffilmark{4,13},
G.W.~Christie\altaffilmark{1,14},
K.~Horne\altaffilmark{3,5,15},
J.~Anderson\altaffilmark{16},
K.Z.~Stanek\altaffilmark{1,6}\\
and\\
D.L.~DePoy\altaffilmark{6},
C.~Han\altaffilmark{17},
J.~McCormick\altaffilmark{18},
B.-G.~Park\altaffilmark{19},
R.W.~Pogge\altaffilmark{6},
S.D.~Poindexter\altaffilmark{6}\\
(The $\mu$FUN Collaboration),\\
I.~Soszy{\'n}ski\altaffilmark{7,20},
M.K.~Szyma{\'n}ski\altaffilmark{7},
M.~Kubiak\altaffilmark{7},
G.~Pietrzy{\'n}ski\altaffilmark{7,20},
O.~Szewczyk\altaffilmark{7},
{\L}.~Wyrzykowski\altaffilmark{7,21},
{K}.~Ulaczyk\altaffilmark{7},
B.~Paczy{\'n}ski\altaffilmark{22}\\
(The OGLE Collaboration)\\
D.M.~Bramich\altaffilmark{5,21},
C.~Snodgrass\altaffilmark{23},
I.A.~Steele\altaffilmark{24},
M.J.~Burgdorf\altaffilmark{24},
M.F.~Bode\altaffilmark{24}\\
(The RoboNet Collaboration)\\
C.S.~Botzler\altaffilmark{4,12},
S.~Mao\altaffilmark{10},
S.C.~Swaving\altaffilmark{4,12}\\
}
\altaffiltext{1}
{Microlensing Follow Up Network ($\mu$FUN)}
\altaffiltext{2}
{Optical Gravitational Lens Experiment (OGLE)}
\altaffiltext{3}
{RoboNet Collaboration}
\altaffiltext{4}
{Microlensing Observations for Astrophysics (MOA) Collaboration}
\altaffiltext{5}
{Probing Lensing Anomalies NETwork (PLANET) Collaboration}
\altaffiltext{6}
{Department of Astronomy, Ohio State University,
140 W.\ 18th Ave., Columbus, OH 43210, USA; 
deokkeun,depoy,dong,gould,pogge,kstanek,sdp@astronomy.ohio-state.edu}
\altaffiltext{7}
{Warsaw University Observatory, Al.~Ujazdowskie~4, 00-478~Warszawa, Poland; 
udalski,soszynsk,msz,mk,pietrzyn,szewczyk,wyrzykow,kulaczyk@astrouw.edu.pl}
\altaffiltext{8}
{Department of Physics, Notre Dame University, Notre Dame, IN 46556, USA;
bennett@nd.edu}
\altaffiltext{9}
{Department of Physics, Astronomy and Materials Science, Missouri
State University, 901 S. National, Springfield, MO 65897 USA;
ayz989f@missouristate.edu}
\altaffiltext{10}
{Jodrell Bank Observatory, The University of Manchester,
Macclesfield, Cheshire SK11 9DL, United Kingdom;
njr,smao@jb.man.ac.uk}
\altaffiltext{11}
{Harvard-Smithsonian Center for Astrophysics, 60
Garden Street, Cambridge, MA 02138; sgaudi@cfa.harvard.edu}
\altaffiltext{12}
{Department of Physics, University of Auckland, 
Private Bag 92019, Auckland, New Zealand;
c.botzler,s.swaving,p.yock@auckland.ac.nz}
\altaffiltext{13}
{Institute of Information and Mathematical Sciences, Massey University,
Private Bag 102-904, North Shore Mail Centre, Auckland, New Zealand;
i.a.bond@massey.ac.nz}
\altaffiltext{14}
{Auckland Observatory, Auckland, New Zealand, gwchristie@christie.org.nz}
\altaffiltext{15}
{SUPA, Physics \& Astronomy, North Haugh,
 St Andrews, KY16~9SS, UK; kdh1@st-andrews.ac.uk}
\altaffiltext{16}
{Dept.\ of Physics and Astronomy, MS-108, Rice University, 
6100 Main Street, Houston, TX-77005, USA;
jay@eeyore.rice.edu}
\altaffiltext{17}
{Department of Physics, Institute for Basic Science Research,
Chungbuk National University, Chongju 361-763, Korea;
cheongho@astroph.chungbuk.ac.kr}
\altaffiltext{18}
{Farm Cove Observatory, Centre for Backyard Astrophysics,
Pakuranga, Auckland New Zealand; farmcoveobs@xtra.co.nz}
\altaffiltext{19}
{Korea Astronomy and
Space Science Institute, Daejon 305-348, Korea; bgpark@kasi.re.kr}
\altaffiltext{20}{Universidad de Concepci{\'o}n, Departamento de Fisica,
Casilla 160--C, Concepci{\'o}n, Chile}
\altaffiltext{21} {Institute of Astronomy  Cambridge University,
Madingley Rd., CB3 0HA Cambridge, UK dmb7@ast.cam.ac.uk}
\altaffiltext{22}{Princeton University Observatory, Princeton, NJ 08544, 
USA; bp@astro.princeton.edu}
\altaffiltext{23}
{Astrophysics Research Centre, School of Physics, Queen's University Belfast,
Belfast BT7 1NN, UK; c.snodgrass@qub.ac.uk}
\altaffiltext{24}
{Astrophysics Research Institute, Liverpool John Moores University,
Twelve Quays House, Egerton Wharf, Birkenhead CH41 1LD, UK;
ias,mfb,mjb@staru1.livjm.ac.uk}

\begin{abstract}
We detect a Neptune mass-ratio ($q\simeq 8\times 10^{-5}$) planetary
companion to the lens star in the extremely high-magnification 
($A\sim 800$) microlensing event OGLE-2005-BLG-169.  If the parent is
a main-sequence star, it has mass $M\sim 0.5\,M_\odot$ 
implying a planet mass of $\sim 13\, M_\oplus$ and projected separation of 
$\sim 2.7\,\au$.  When intensely monitored over their peak, 
high-magnification events similar to OGLE-2005-BLG-169 have nearly
complete sensitivity to Neptune mass-ratio planets with projected separations of
0.6 to 1.6 Einstein radii, corresponding to 1.6--4.3 AU in the present case.
Only two other such events were monitored
well enough to detect Neptunes, and so this detection by itself suggests
that Neptune mass-ratio planets are common.  Moreover,
another Neptune was recently discovered at a similar distance
from its parent star in a low-magnification event, which are more common but are
individually much less sensitive to planets.  Combining the two detections
yields 90\% upper and lower frequency
limits $f=0.37^{+0.30}_{-0.21}$ over just 0.4 decades of planet-star separation.
In particular, $f>16\%$ at 90\% confidence.
The parent star hosts no
Jupiter-mass companions with projected separations within a factor 5
of that of the detected planet.  The lens-source relative
proper motion is $\mu\sim 7$--$10\,\masyr$, implying that if the lens is 
sufficiently
bright, $I\la 23.8$, it will be detectable by the {\it Hubble Space Telescope}
by 3 years after peak.  This would permit a more precise 
estimate of the lens mass and distance, and so the mass and projected separation
of the planet.  Analogs of OGLE-2005-BLG-169Lb
orbiting nearby stars would be difficult to detect by other
methods of planet detection, including radial velocities, transits, or
astrometry.

\end{abstract}


\section{Introduction
\label{sec:intro}}

The regions of our Solar System beyond Mars contain giant planets,
as well as asteroids and comets, which are believed to be remnants of the
process that formed these bodies, and also moons, which may serve
as analogs to bodies involved in the late stages of planet formation.
However, despite some 170 planet discoveries over the past decade,
the analogous regions around other mature stars remain relatively inaccessible
to us.  As radial velocity (RV) survey time baselines have grown, they
have begun to detect gas giants in these regions, but RV is sensitive
to Neptune-mass planets only when they are much closer to their parent
stars.  Transit surveys are even more heavily biased toward close-in
planets.  Astrometric sensitivity does peak at large
orbits but is fundamentally restricted to orbital periods that are shorter
than the survey.

By contrast, microlensing sensitivity peaks at the Einstein ring, which
is typically at 2--4 AU,
depending on the mass and distance of the
host star, and it extends several times farther out.  Moreover microlensing
detections are ``instantaneous snapshots'' of the system, not
requiring an orbital period to elapse.  Hence, microlensing
can potentially yield important information about the
intermediate-to-outer regions of extrasolar planetary
systems that are difficult to probe by other techniques.

Gravitational microlensing occurs when a ``lens'' star becomes closely aligned,
within an angular Einstein radius $\theta_\e$, with a more distant source
star.  The source is magnified by an amount that grows monotonically
as it approaches the lens, and diverges inversely with separation
for extremely close encounters \citep{einstein36,pac86}.

Planets hosted by the lens star induce two, generally distinct, perturbations
on the single-lens magnification pattern: a small ``planetary caustic''
(closed contour
of formally infinite magnification) directly associated
with the planet and an even smaller ``central caustic'' closely aligned with
the host.  All microlensing events have the potential
to probe the planetary caustic, but owing to its small extent and
the random trajectory of the source, the probability for the source
to encounter the perturbed region is low.

By contrast, the small fraction of events that reach very high magnification
(and so small lens-source separation)
{\it automatically} probe the region of the central caustic
\citep{griestsafi,rhie00,bond02,rattenbury02}.
Hence, even low-mass
planets lying anywhere sufficiently near the Einstein ring are
virtually guaranteed to perturb the light curve \citep{science}.
Since the peak of the event can, at least in principle, be predicted
in advance, it is possible to focus limited observing resources to make
the intensive observations over the peak that are required to
detect and characterize the relatively subtle signal.

The Microlensing Follow Up
Network\footnote{http://www.astronomy.ohio-state.edu/$\sim$microfun/}
 ($\mu$FUN) has adopted a strategy of trying to recognize the few
prospective high-magnification events among
the roughly 500 microlensing events annually alerted by the
Optical Gravitational Lens
Experiment\footnote{http://www.astrouw.edu.pl/$\sim$~ogle/ogle3/ews/ews.html}
(OGLE-III) Early Warning System \citep{ews} and the roughly 50 events
annually alerted by the Microlensing Observations for
Astrophysics\footnote{http://www.massey.ac.nz/$\sim$iabond/alert/alert.html}
collaboration (MOA).
During the 2005 season, this strategy led to the discovery of 2 planets.
The first was the Jovian mass-ratio OGLE-2005-BLG-071Lb
\citep{ob05071}.  Here we report the
second such planet, a Neptune-mass-ratio companion to the
lens in OGLE-2005-BLG-169.

\section{Event Discovery and Data
\label{sec:discovery}}

On 2005 Apr 21, the OGLE collaboration alerted OGLE-2005-BLG-169
as probable microlensing of a faint $(I=19.4)$ source toward
the Galactic bulge, using
the 1.3m Warsaw telescope in Chile
(operated by the Carnegie Institution of Washington).  After
observations by OGLE and the 1.3m $\mu$FUN SMARTS telescope in Chile
showed the event to be extremely high magnification, 
the observers at the 2.4m $\mu$FUN MDM telescope in Arizona interrupted their
regular program to obtain more than 1000 exposures over the peak.
Additional
data come from the 0.35m $\mu$FUN Nustrini telescope in Auckland, New Zealand
and the 2.0m PLANET/RoboNet\footnote{http://planet.iap.fr/,
http://www.astro.livjm.ac.uk/RoboNet} Faulkes Telescope North
in Hawaii.  We analyze a total of (340,22,1025,74,31) images
with typical exposure times of (120,300,10,120,100) seconds in the
$(I,I,I,{\rm clear},R)$ passbands, respectively, from these 5 telescopes.
In addition, 6 $V$ band images from $\mu$FUN SMARTS permit determination
of the source color.

\begin{figure}
\plotone{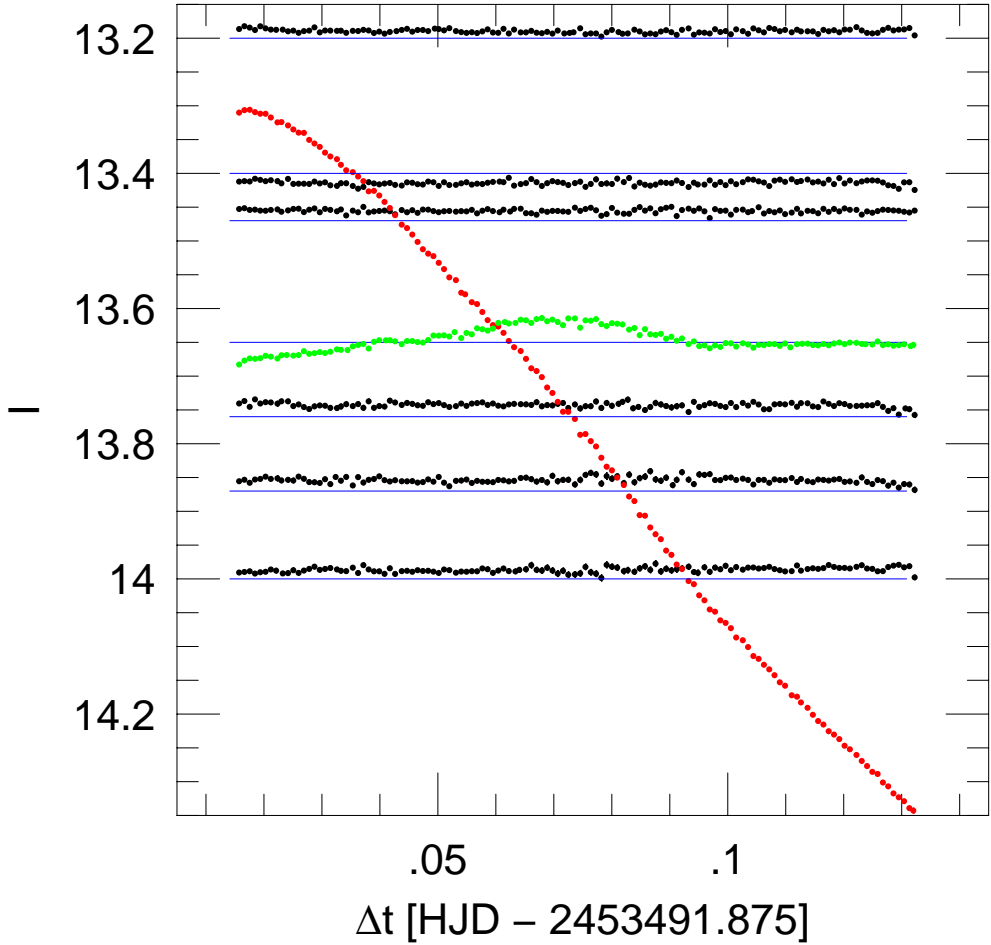}
\caption{
Light curves of 6 reference stars
({\it black}) compared to
microlensed source ({\it red}) as derived by the DIA reductions
of the MDM data.  The deviation of this curve from a single-lens
microlensing fit (derived from the non-MDM points) is shown in {\it green}.
Horizontal {\it blue} lines are shown to aid in judging the constancy
of the various stars.
}
\label{fig:comp_stars}
\end{figure}

All data
were reduced using the OGLE data pipeline based on difference imaging analysis
(DIA) \citep{wozniak}.  The $\mu$FUN MDM data were also reduced
using the ISIS pipeline \citep{alard98,alard00,hartman04}. To test
for any systematics in this crucial data set, we report results below
derived from these two completely independent pipelines.  See also 
Figure~\ref{fig:comp_stars}.

\section{Light Curve Model
\label{sec:lc}}

Planets are discovered in microlensing events from the brief perturbation
they induce on a single-lens light curve \citep{mao91,gouldloeb92}:
$F(t) = F_{\rm s} A[u(t)] + F_{\rm b}$, where
$A(u) = (u^2+2)/u(u^2+4)^{1/2}$,
$u(t) = (\tau^2+u_0^2)^{1/2}$,
$\tau = (t-t_0)/t_\e$,
Here, $F(t)$ is the observed flux, $F_{\rm s}$ is the source flux,
$F_{\rm b}$ is flux due to any unlensed background light, $A$ is the
magnification, and ${\bf u}=(\tau,u_0)$ is the vector position of the
source normalized to $(\theta_\e)$,
expressed in terms of the three geometrical parameters of the
event: the time of closest approach $t_0$, the normalized impact
parameter $u_0$, and the Einstein timescale $t_\e$.

To describe the planetary perturbation,
3 ``binary-lens'' parameters are required
in addition to the 3 single-lens geometric parameters
$(t_0,u_0,t_\e)$.  These are the
binary mass ratio, $q$, the separation of the components (in units
of $\theta_\e$), $b$, and angle of the source trajectory
relative to the binary axis, $\alpha$.
Finally, a seventh parameter, $\rho\equiv \theta_*/\theta_\e$, is required
whenever the angular radius of the source $\theta_*$  plays a significant
role, in particular, whenever the source crosses a caustic.

For planetary lenses with caustic crossings, there are 7
directly observable and pronounced light-curve features that directly
constrain
the 7 model parameters, up to a well-understood two-fold degeneracy
 \citep{dominik99} that takes $b\leftrightarrow b^{-1}$.
Three of these, the epoch, duration, and height of the primary lensing event,
strongly constrain $(t_0,t_\e,u_0)$.  The remaining four,
the two caustic-crossing times (entrance and exit) and
the height and duration of one of the caustic crossings, then constrain
$(b,q,\alpha,\rho)$.

As we now show, OGLE-2005-BLG-169 is indeed a caustic-crossing event.
However, only the caustic exit (but not entrance) was well-resolved.
This leads to a 1-dimensional degeneracy among the model parameters,
which is then partially broken by secondary, less pronounced, features
of the light curve.

\begin{figure}
\plotone{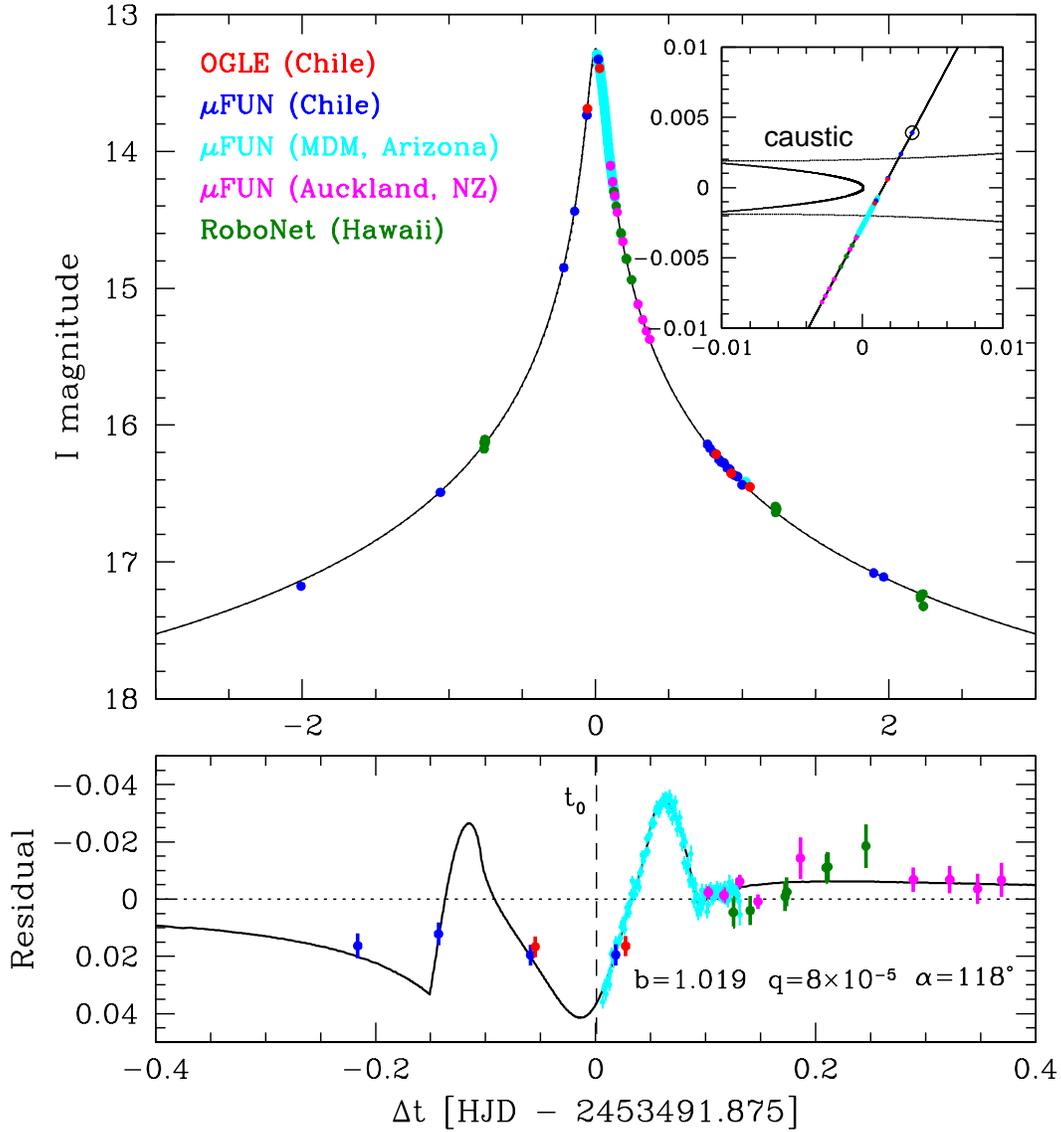}
\caption{
Data and best-fit model of OGLE-2005-BLG-169.
Lower panel shows difference between this model
and a single-lens model with the same ($t_0,u_0,t_\e,\rho$).
It displays
the classical form of a caustic entrance/exit that is
often seen in binary microlensing events, where the amplitudes
and timescales
are several orders of magnitude larger than seen here.  MDM data trace
the characteristic slope change at the caustic exit ($\Delta t=0.092$)
extremely well, while the entrance is tracked by a single point
($\Delta t=-0.1427$).  The dashed line indicates the time $t_0$.
Inset shows source path through the
caustic geometry and indicates the source size, $\rho$.
}
\label{fig:lc}
\end{figure}

The residuals to the single-lens models shown in Figure~\ref{fig:lc}
exhibit a kink in slope at $\Delta t=0.092\,$day, where
$\Delta t$ is the time elapsed since HJD 2453491.875 (2005 May 1 09:00).
The kink is equally apparent in both sets of MDM reductions.
While the {\it incident slope} of this kink in the light curve depends
on the particular single-lens model used, the {\it slope discontinuity}
is model-independent.
Such a change in slope is induced by a caustic exit, when
the trailing limb of the source crosses the caustic, causing
two of its images to merge and finally vanish.

\begin{figure}
\plotone{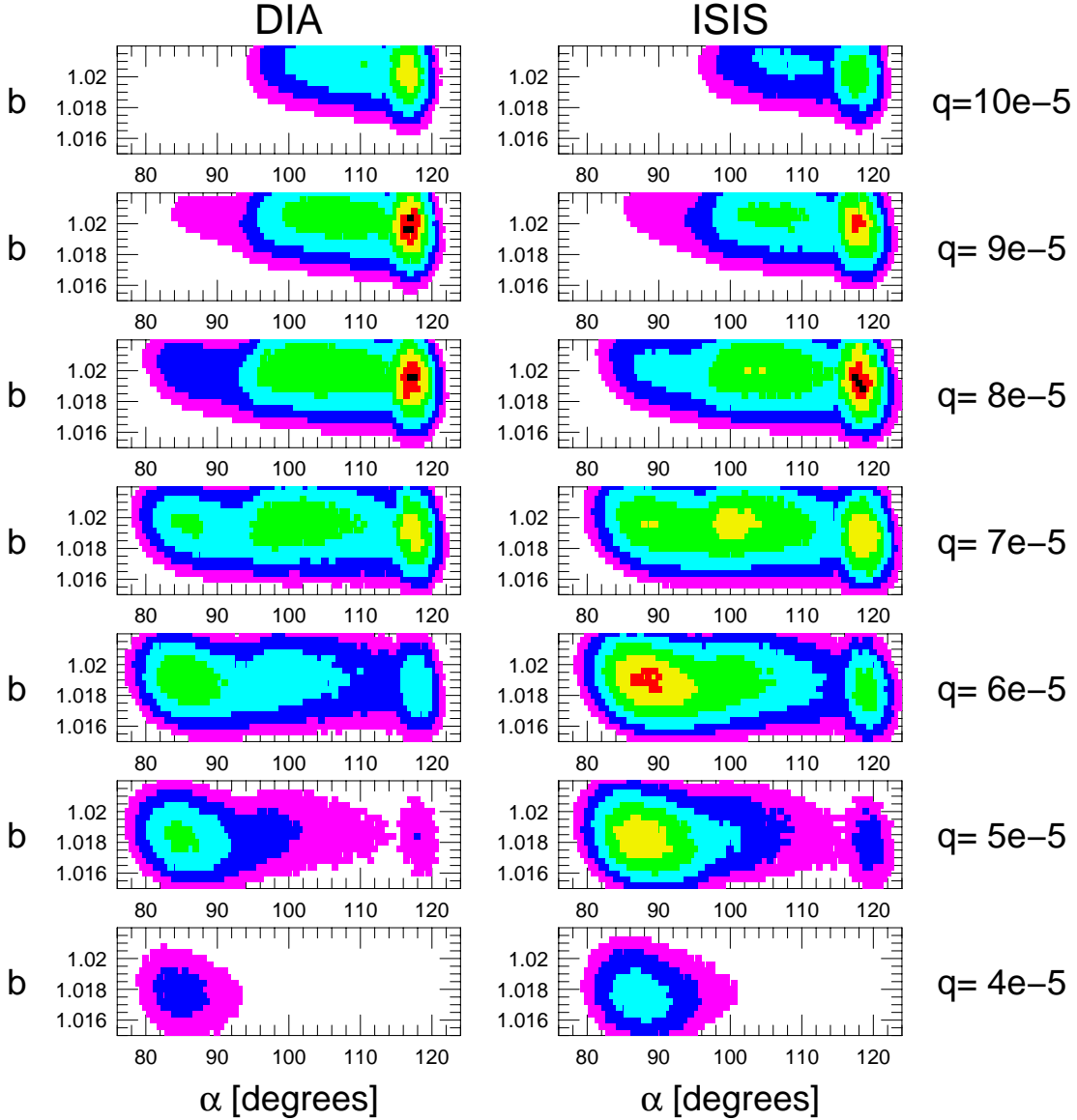}
\caption{
 $\Delta\chi^2$ contours relative to the respective minima
for light-curve fits using the DIA and ISIS
reductions of the MDM data.  $\sigma\equiv \sqrt{\Delta\chi^2}<1,2,3,4,5,6,7$
are shown by black, red, yellow, green, cyan, blue, and magenta.
Note that abscissae are compressed, so the $6\,\sigma$ contours have
an axis ratio of approximately 100 (with $\alpha$ expressed in radians),
which reflects the 1-D degeneracy
discussed in the text.  Both reductions have their
minimum at $(\alpha,q)\sim (120^\circ,8\times 10^{-5})$.
Similar contours for $b < 1$
yield additional solutions that are included in Table 1.
The full $3\,\sigma$ mass-ratio range is confined
to 5--10$\times 10^{-5}$ even allowing for both reductions.
}
\label{fig:dia_isis}
\end{figure}

To extract model parameters, we undertake
a brute-force search of parameter space.
We hold the three parameters $(b,\alpha,q)$ fixed at a grid of values
while minimizing $\chi^2$ over the remaining four parameters,
using the values of $(t_0,u_0,t_\e)$ derived from the overall shape
of the light curve as seeds.  We find two distinct local minima that
obey the $b\leftrightarrow b^{-1}$ degeneracy.
These minima are embedded in elongated $\Delta\chi^2$ valleys 
(Fig.~\ref{fig:dia_isis}),
with [$\alpha$(in radians)$:b$] axis ratio $\sim 100$, which occur because there
are only 6 pronounced features in the light curve to constrain 7 model
parameters.  However, while the caustic entrance is not
well resolved, the point at $\Delta t = -0.1427\,$day
does lie on this entrance and thereby singles out the solutions
shown in Table~\ref{tab:models}.  Nevertheless, to be conservative,
we quote $3\,\sigma$ errors for light-curve parameters, which take account of
both the elongated valleys and the two reductions.  Most importantly,
\begin{equation}
q = 8^{+2}_{-3}\times 10^{-5},\qquad b=1.00\pm 0.02\qquad (3\,\sigma).
\label{eqn:qandb}
\end{equation}
\section{Mass of the Host Star and Planet
\label{sec:constraints}}

In microlensing events, the physical parameters (lens mass $M$,
lens-source relative parallax $\pi_\rel$,
and relative proper motion $\mu$) are related
to the ``observable'' event parameters ($t_\e$, $\theta_\e$, and
the microlens parallax \citep{gould04}, $\pi_\e$), by
\begin{equation}
\theta_\e = \sqrt{\kappa M \pi_\rel},
\quad
\pi_\e = \sqrt{\kappa M \over \pi_\rel},
\quad
t_\e = {\theta_\e\over\mu},
\label{eqn:parms}
\end{equation}
where $\kappa\equiv 4G/(c^2\au)$.  Of these,
only $t_\e$ is routinely measurable.
However, in caustic-crossing events, one can usually also measure
$\theta_\e$, which both directly constrains the mass-distance relation
and yields a measurement of $\mu$.
We first determine the angular source radius $\theta_*$ using the
standard approach \citep{ob03262}, finding
$\theta_*=0.44\pm 0.04\,\muas$.
Together with the parameter measurements
$\rho=4.4^{+0.9}_{-0.6}\times 10^{-4}$ and $t_\e=43\pm 4\,$days, this yields
$3\,\sigma$ ranges,
\begin{equation}
\theta_\e = {\theta_*\over\rho} = 1.00\pm 0.22\,\mas,
\quad
\mu = {\theta_\e\over t_\e} = 8.4 \pm 1.7\,\masyr.
\label{eqn:thetaemu}
\end{equation}

Another constraint comes from the upper limit on the lens flux,
which cannot exceed the background flux measurement,
$F_{\rm b}$ (corresponding to $I_{\rm b}=19.8$).
We also derive from the light curve a weak constraint on the microlens
parallax, $\pi_{\e,\parallel} = -0.086 \pm 0.261$,
which we include for completeness.  Here,
$\pi_{\e,\parallel} \equiv \pi_\e\cos\psi$ and $\psi$ is the angle
between the direction of lens-source relative motion and the position
of the Sun at $t_0$ projected on the plane of the sky \citep{gould04}.

These measurements, together with a Bayesian prior for the lens
distances and masses from a \citet{han03} Galaxy model and
\citet{gould00} mass function (but with a Salpeter [$-2.35$] slope
at the high end), together with the assumption that all stellar objects 
along the line of sight are equally likely to harbor planets, 
yield a probability distribution
for the lens mass.  We find that the probabilities that the lens
is a main-sequence star (MS), white dwarf (WD), neutron star (NS),
and black hole (BH) are
respectively 55\%, 32\%, 11\%, and 2\% .
However, Neptune mass-ratio
planets around NSs must be quite rare because the $\sim 0.2$
second oscillations they induce would easily show up in pulsar timing
residuals.  Since BHs and NSs form by a similar process, this may
also argue against BHs as the host.  MSs and WDs represent different
life stages of the same class of stars.
If the host is a MS, then the
median and 90\% confidence interval for the mass and distance are given by
\begin{equation}
M=0.49^{+0.23}_{-0.29}\,M_\odot,\qquad 
D_L=2.7^{+1.6}_{-1.3}\,\kpc, 
\qquad (90\%\,\rm confidence)
\label{eqn:masslens}
\end{equation}
implying that the best estimates for the planet mass and separation are
$m_p = q M \sim 13\,M_\oplus$ and
$r_\perp = b\theta_\e D_{\rm L} \sim 2.7\,\au$, where
$D_{\rm L}$ is the distance to the lens.
Since WD masses are sharply peaked
at $M\sim 0.6\,M_\odot$, the corresponding planet characteristics
lie near the center of the same range.  Future observations by the
{\it Hubble Space Telescope} could distinguish between MS and WD hosts,
as well as measuring the mass and distance of the former 
provided that the lens is brighter than $I\sim 23.8$ \citep{bennett06}.

\section{Limits on other Companions
\label{sec:limits}}

Does lens OGLE-2005-BLG-169L have planets other than OGLE-2005-BLG-169Lb?
Central-caustic events allow one to address this question because all the
companions to the lens star perturb the central caustic
\citep{gaudi98}.
Moreover, the combined perturbation from two
planets is very nearly the sum of the separate perturbations, unless
the two planets are closely aligned \citep{rattenbury02,han05}.
This allows us to subtract
the perturbation from the one detected planet and apply the same
search technique to the resulting (nearly single-lens) light curve
to look for others.  We find none, and so place upper limits
on the presence of other planets.
In particular, we exclude Jupiter-mass
$(q=2\times 10^{-3})$ planets at projected separations within a factor 5.5
of the Einstein radius
($0.18<b<5.5$) and Saturn-mass planets within a factor 3.5.

\section{Discussion
\label{sec:discuss}}

\subsection{Low-mass Planets Are Common at Several AU
\label{sec:common}}

OGLE-2005-BLG-169 is one of only three high-magnification events
with sensitivity to the central caustics induced by cold Neptunes.
It and one other event,
MOA-2003-BLG-32/OGLE-2003-BLG-219 \citep{science}, were sensitive to
$q=8\times 10^{-5}$ planets throughout the ``lensing zone'' ($0.6<b<1.6$),
while OGLE-2004-BLG-343 \citep{ob04343} was sensitive over 40\% of the
lensing zone.  This implies $\langle f\rangle=1/2.4=42\%$
for the expected fraction
of stars hosting cold Neptunes within the lensing zone (0.4 decades of
projected separation), but with very large uncertainty due to small
number statistics.

However, we can improve our estimate by incorporating the
detection of OGLE-2005-BLG-390Lb, which is a very similar
$(b=1.6,q=7.6\times 10^{-5})$ planet that was detected through the other
(planetary-caustic) microlensing channel \citep{ob05390}.
The expected number of detections
of $q=8\times 10^{-5}$ lensing-zone planets (through this planetary-caustic
channel) was $1.75 f$ for 1995-1999 \citep{gaudi02}.
We estimate that the expected number
for 2000-2005 is a factor 1.5 higher based on a comparison of the
survey characteristics during these two periods.  Hence, the single
detection through this channel yields $\langle f\rangle=23\%$, also with large errors.
If we combine the two channels, and impose a uniform prior, we find a
median and 90\% upper and lower limits of $f=0.37^{+0.30}_{-0.21}$, in
particular, a 90\% confidence lower limit of $f>16\%$.

\subsection{Unique capabilities of microlensing
\label{sec:unique}}

\begin{figure}
\plotone{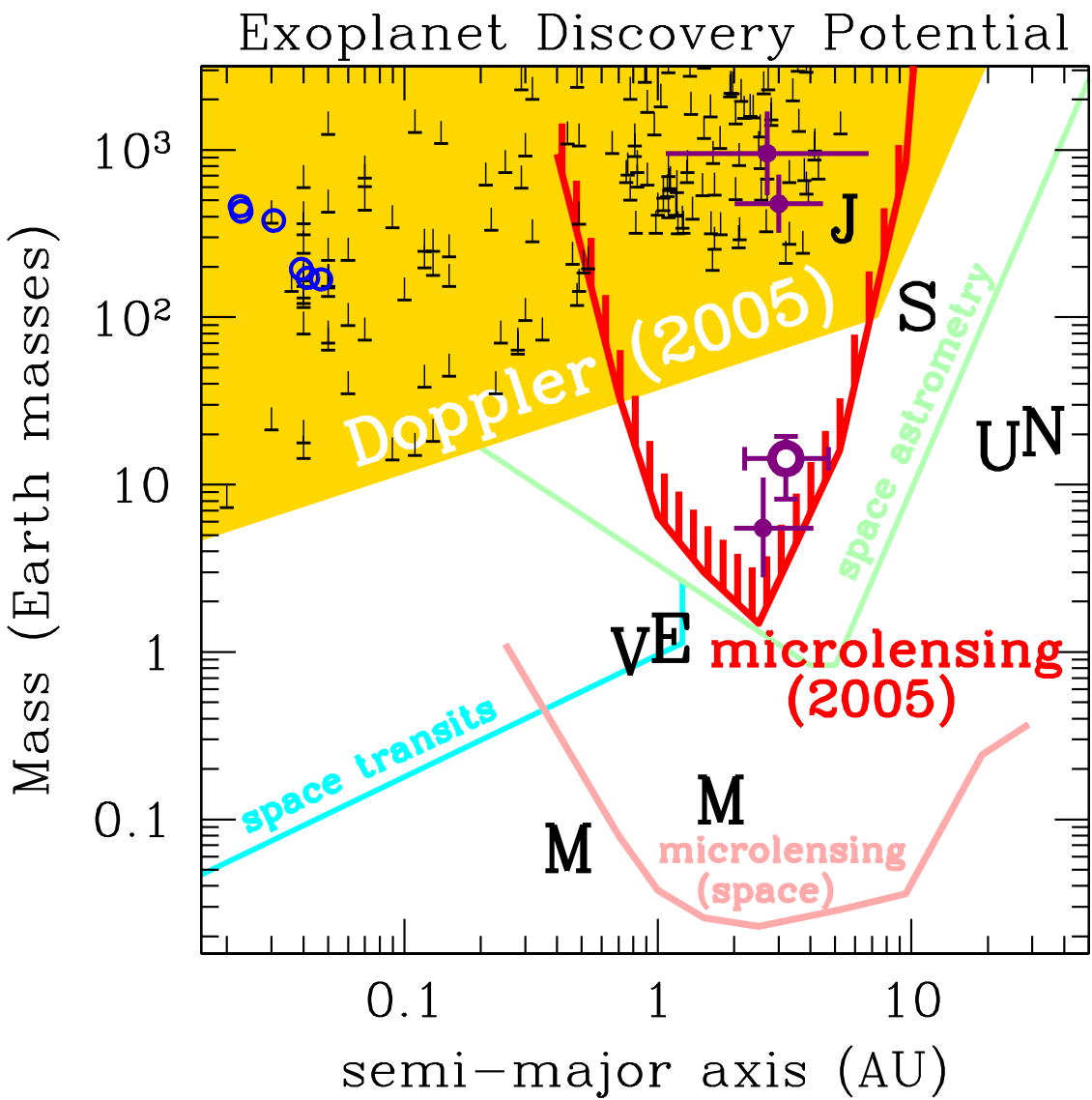}
\caption{
Exoplanet discovery potential and detections
are shown as functions of planet mass and semimajor axis.  Potential
is shown for current ground-based RV ({\it yellow}) and, very approximately,
microlensing ({\it red}) experiments as well as
future space-based transit ({\it cyan}), astrometric ({\it green}), and
microlensing ({\it peach}) missions.  Planets {\it discovered} using the
transit ({\it blue}), RV ({\it black}), and microlensing ({\it magenta})
techniques are shown as individual points,
with OGLE-2005-BLG-169Lb displayed as an open symbol.
Solar system planets are indicated by their initials for comparison.
}
\label{fig:m_vs_sep}
\end{figure}

Can analogs to OGLE-2005-BLG-169Lb (with the same mass ratio and semi-major
axis) be detected by other techniques, such as RV, 
transits, and astrometry?  To make a strict comparison, it would
be necessary to know three additional parameters of the event, namely the
lens and source distances, $D_{\rm L}$ and $D_{\rm S}$,
and the projection angle $\psi$ of its semi-major
axis $a$ to the line of sight, i.e., $\sin\psi=r_\perp/a$ where $r_\perp$
is the projected star-planet separation.  The stellar mass and semi-major axis
are then related to the event parameters by
$M = \theta_\e^2/\kappa\pi_\rel$
and
$a = D_{\rm L}\theta_\e b \csc\psi$.
Hence, assuming circular orbits,
the velocity amplitude $v\sin i$ and orbital period $P$ would be
\begin{equation}
v\sin i = {q c\sin i\over 2}\sqrt{\theta_\e\sin\psi\over b(1-x)}
\rightarrow 0.85\,{\rm m\,s^{-1}}\sqrt{\sin\psi\over (1-x)}\sin i,
\label{eqn:vsini}
\end{equation}
\begin{equation}
P = {4\pi D_{\rm L}\over c}\sqrt{b^3\csc^3\psi (1-x)\theta_\e}
\rightarrow 9.4\,{\rm yrs} 
{D_{\rm L}\over 2.7\,{\rm kpc}}\sqrt{1-x\over 0.66}
\biggl({\csc\psi\over 1.3}\biggr)^{3/2}.
\label{eqn:period}
\end{equation}
Here $x\equiv D_{\rm L}/D_{\rm S}$ and the evaluations are for 
$\theta_\e=1\,\mas$,
$q=8\times 10^{-5}$
and $b=1$.
These equations imply that RV detection of OGLE-05-BLG-169Lb analogs would be
extremely difficult both because of the low velocity amplitude and
the long period.  Similarly, the long period renders transit detection
impossible unless a transit experiment of considerably longer duration
than {\it Kepler} were organized.  
Finally, 
the amplitude of astrometric motion would be
\begin{equation}
\alpha
= q {D_{\rm L}\over D_{\rm analog}}
\theta_\e b \csc\psi,
\rightarrow 
28\,\mu{\rm as}\,{D_{\rm L}\over 2.7\,\kpc}\, {10\,\pc\over D_{\rm analog}}\,
{\csc\psi\over 1.3},
\label{eqn:astrometry}
\end{equation}
where $D_{\rm analog}$ is the distance to the local analog system.
This amplitude would be easily detectable by the {\it Space Interferometry
Mission (SIM)}, although at the typical periods, this would require an
extended {\it SIM} mission.  See Figure~\ref{fig:m_vs_sep}.

\subsection{Neptunes or Failed-Jupiters?
\label{sec:nature}}

Both OGLE-2005-BLG-169Lb and OGLE-2005-BLG-390Lb are in the
cold outer regions of their planetary systems, with expected surface
temperatures of $\sim 70\,$K and $\sim 50\,$K, corresponding roughly to
the environments of Saturn and Neptune, respectively.  Both
have Neptune/Sun mass ratios, with absolute mass estimates of
$13\,M_\oplus$ and $6\,M_\oplus$, respectively.  They must have
a large fraction of rock and ice, but whether these are covered
with a thick coat of gas, like Uranus and Neptune, or whether
they are ``naked'' super-Earths such as are theorized to
have formed the cores of Jupiter and Saturn, is unclear.
If such cores formed routinely but usually failed to accrete
the ambient gas before it dispersed, this would account for
the high frequency of these objects \citep{ida05}.  Moreover, the absence of
gas giants within a factor 5.5 of the OGLE-2005-BLG-169's Einstein ring
is consistent with the idea that the detected planet is such
a ``failed Jupiter'' \citep{laughlin05}.
One could gain further clues
by mapping out the mass and separation distributions
of a larger sample.

\acknowledgments

We acknowledge the following support:
NSF AST-042758 (AG,SD); NASA NNG04GL51G (DD,AG,RP);
Polish MEiN 2P03D02124, NSF AST-0204908, NASA grant NAG5-12212 (OGLE);
Polish FNP SP13/2003 (AU); NSF AST-0206189, NASA NAF5-13042 (DPB);
NSF AST-007480 (A-YZ); Menzel Fellowship Harvard College Obs (BSG);
SRC Korea Science \& Engineering Foundation (CH);
Korea Astronomy \& Space Science Institute (B-GP);
Marsden Fund of NZ (IAB,PCMY);
Deutsche Forschungsgemeinschaft (CSB);
PPARC, EU FP6 programme ``ANGLES'' ({\L}W,SM,NJR);
PPARC (RoboNet);
Dill Faulkes Educational Trust (Faulkes Telescope North);
Assistance by Lydia Philpott, Jan Snigula and the Computer Science
Department of the University of Auckland is acknowledged.
We thank the MDM staff for their support.
All findings are those of the authors
and do not necessarily reflect NSF views.

\begin{deluxetable}{l r r r r r r r r r r r c}
\tablecaption{\label{tab:models} Light Curve Models}
\tablewidth{0pt} \tablehead{ 
\colhead{Pipe-} & 
\colhead{$\Delta\chi^2$} &
\colhead{$t_0-t_{\rm ref}$} & 
\colhead{$u_0$} & 
\colhead{$t_\e$} &
\colhead{$b$} & 
\colhead{$q$} &
\colhead{$\alpha$} &
\colhead{$\rho$} &
\colhead{$\theta_\e$} &
\colhead{$\mu$}
\\
\colhead{line} &
\colhead{} &
\colhead{day} &
\colhead{$\times 10^3$} &
\colhead{day} &
\colhead{} &
\colhead{$\times 10^5$} &
\colhead{deg} &
\colhead{$\times 10^4$} &
\colhead{mas} &
\colhead{mas/yr}
}
\startdata 

DIA  &  0.00 & 0.0008 & 1.24 & 42.27 & 1.0198 & 8.6 & 117.0 &  4.4 & 1.00 &  8.6\\
ISIS &  0.00 & 0.0009 & 1.23 & 42.56 & 1.0194 & 8.2 & 118.2 &  4.7 & 0.93 &  8.0\\
DIA  &  0.27 & 0.0004 & 1.25 & 42.09 & 0.9819 & 8.3 & 122.6 &  3.9 & 1.12 &  9.7\\
ISIS &  2.33 & 0.0007 & 1.17 & 44.69 & 0.9825 & 7.3 & 123.5 &  4.0 & 1.11 &  9.0\\
\enddata
\tablecomments{$t_{\rm ref}=$ HJD 2453491.875 (2005 May 1 09:00).}
\end{deluxetable}

\end{document}